\documentstyle[aps,epsfig,amsfonts,prb]{revtex}

\begin{document}

\def\simgt{\mathrel{\setbox0\hbox{$>$}\setbox1\hbox{$\sim$}\ifdim\wd0>\wd1
\hbox to0pt{\hskip0.5\wd0\raise-.6ex\hbox
to0pt{\hss\copy1\hss}\hss}\raise.4ex\box0
\else
\hbox to0pt{\hskip0.5\wd1\raise.4ex\hbox
to0pt{\hss\copy0\hss}\hss}\raise-.6ex\box1
\fi}}
\def\simlt{\mathrel{\setbox0\hbox{$<$}\setbox1\hbox{$\sim$}\ifdim\wd0>\wd1
\hbox to0pt{\hskip0.5\wd0\raise-.6ex\hbox
to0pt{\hss\copy1\hss}\hss}\raise.4ex\box0
\else
\hbox to0pt{\hskip0.5\wd1\raise.4ex\hbox
to0pt{\hss\copy0\hss}\hss}\raise-.6ex\box1
\fi}}

\def\kB{k_B}    
\def\Ref{^{\rm ref}} 
\def\withQ#1{^{(#1)}} 
\def\Q{\withQ Q}
\def\withQo{\withQ{Q_0}}

\def\A{{\cal A}}         
\def\G{{\cal G}}         
\def\C{{\cal C}}         
\def\c{ c_2}             
\def\K{{\cal K}}         
\def\g{\gamma}           
\def\eps{\varepsilon}    
\def\epsbar{\bar\eps}    
\def\u{\tilde u}           
\def\epsNum{\epsilon_{\tt\#}}

\def\PDE{{\sc pde}}
\def\ODE{{\sc ode}}
\def\HRT{{\sc hrt}}
\def\LOGA{{\sc loga}}
\def\ORPA{{\sc orpa}}
\def\OZ{{\sc oz}}
\def\HNC{{\sc hnc}}
\def\RPA{{\sc rpa}}
\def\RG{{\sc rg}}
\def\MC{{\sc mc}}
\def\CPU{{\sc cpu}}

\let\rho\varrho

\title{The Hierarchical Reference Theory\\
as applied to square well fluids of variable range}

\author{Albert~Reiner$^*$ and Gerhard~Kahl}

\address{Institut f\"ur Theoretische Physik and
Center for Computational Materials Science,\\
Technische Universit\"at Wien,
Wiedner Hauptstra\ss e~8--10, A--1040 Vienna, Austria.\\
{}$^*$e-mail: {\tt areiner@tph.tuwien.ac.at}}

\maketitle

\begin{abstract}
Continuing our investigation into the numerical properties of the {\it
Hierarchical Reference Theory}, we study the square well fluid of
range $\lambda$ from slightly above unity up to 3.6. After briefly
touching upon the core condition and the related decoupling assumption
necessary for numerical calculations, we shed some light on the way an
inappropriate choice of the boundary condition imposed at high density
may adversely affect the numerical results; we also discuss the
problem of the partial differential equation becoming stiff for
close-to-critical and sub-critical temperatures.  While agreement of
the theory's predictions with simulational and purely theoretical
studies of the square well system is generally satisfactory for
$\lambda\simgt2$, the combination of stiffness and the closure chosen
is found to render the critical point numerically inaccessible in the
current formulation of the theory for most of the systems with
narrower wells. The mechanism responsible for some deficiencies  is
illuminated at least partially and allows us to conclude that
the specific difficulties encountered for square wells are not likely
to resurface for continuous potentials.\end{abstract}


\section{Introduction}

\label{sec:intro}

In a large part of the density-temperature plane, integral equation
theories are a reliable tool for studying thermodynamic and structural
properties of, among others, simple one-component fluids
\cite{b:allg:1}; unfortunately, in the vicinity of a liquid-vapor
critical point, integral equations are haunted by a host of
difficulties, leading to a variety of shortcomings such as incorrect
and non-matching branches of the binodal, classical values at best for
the critical exponents, or other deviations from the correct behavior
at the critical singularity  \cite{b:hrt:1}. Asymptotically close to the
critical point, on the other hand, renormalization group (\RG) theory is
the
instrument of choice for describing the fluid; in general, however,
\RG\ approaches do not allow one to derive non-universal quantities
from microscopic information only, {\it i.~e.{}}\ from knowledge of the
forces
acting between the fluid's particles alone. One of the theories
devised to bridge the conceptual gap between these complementary
approaches is the {\it Hierarchical Reference Theory} (\HRT) first put
forward by Parola and Reatto
 \cite{b:hrt:1,hrt:1,hrt:2,hrt:11,hrt:10,hrt:3,hrt:4,hrt:12,hrt:5,hrt:6,hrt:8,hrt:7}: In this
theory the introduction of a cut-off wavenumber $Q$ inspired by
momentum space \RG\ theory and, for every value of $Q$, of a
renormalized potential $v\Q(r)$ means that only non-critical systems
have to be considered at any stage of the calculation; consequently,
integral equations may successfully be applied to every system with
$Q>0$, and critical behavior characterized by non-classical critical
exponents is recovered only in the limit $Q\to0$.

While applicability of \HRT\ to a number of interesting systems,
ranging from a lattice gas or Ising model \cite{hrt:6} to various
one-component fluids \cite{hrt:10,hrt:3,hrt:4} even including
three-body interactions \cite{hrt:7,hrt:18}, internal degrees of
freedom \cite{hrt:19}, or non-hard-core reference systems \cite{hrt:17},
was demonstrated early on, the main focus of research on \HRT\ has
since shifted to the richer phase behavior of binary systems
\cite{hrt:17,hrt:9,hrt:13}. Nevertheless, in the light of \HRT's high
promise and low penetration into the liquid physics community, further
study and critical assessment of this theory seem worthwhile, even and
foremost in the case of simple one-component fluids: indeed, it is in
this comparatively simple setting that we may gain important insights
into the numerical side of the theory, and barring special mechanisms
relevant to some specific model system only, any problems uncovered
here must be expected to haunt more advanced applications of \HRT,
too. In our work we have found it convenient to restrict ourselves
even further, implementing \HRT\ in its usual formulation
\cite{b:hrt:1,ar:4,ar:th} for purely pairwise additive interactions
{\it via}\ a potential $v(r)$ obtained from the superposition of an
infinitely repulsive hard sphere serving as reference system,
$v\Ref(r)=v^{{\rm hs}}(r)$, and a predominantly attractive tail $w(r)$,
$\tilde
w(0)<0$. Here we have made use of the notation introduced previously
\cite{ar:4}: superscripts always denote the system a quantitiy refers to
(here, ``ref'' and ``hs'' for the reference system and hard spheres,
respectively; similarly, ``$(Q)$'' for the system with cut-off $Q$),
and a tilde indicates Fourier transformation.  In the present
contribution we apply our recent re-implementation of the theory
\cite{ar:4} to one of the simplest potentials exhibiting phase
separation, {\it viz.{}}\ the square well potential
$v^{{\rm sw}[-\epsilon,\lambda,\sigma]}$ (cf.~sub-section III~B of
ref.~\onlinecite{ar:4}):
\begin{equation}\label{eq:def:potSW}
\begin{array}{c}
v^{{\rm sw}[-\epsilon,\lambda,\sigma]}(r) =
v^{{\rm hs}[\sigma]}(r)+w^{{\rm sw}[-\epsilon,\lambda,\sigma]}(r)
\\
v^{{\rm hs}[\sigma]}(r) = \left\{\begin{array}{ccc}
+\infty &:& r<\sigma \\
0      &:& r>\sigma
\end{array}\right.
\\
w^{{\rm sw}[-\epsilon,\lambda,\sigma]}(r) = \left\{\begin{array}{ccc}
-\epsilon &:& r<\lambda\,\sigma \\
0        &:& r>\lambda\,\sigma\,.
\end{array}\right.
\end{array}
\end{equation}
Considering density-independent potentials only and chosing the hard
core diameter $\sigma$ and the well's depth $\epsilon$ as units of
length and energy, respectively, the attractive well's range $\lambda$
is the sole remaining parameter; in this report we will study values
of $\lambda$ from slightly above unity up to 3.6.

With just one parameter, {\it viz.{}}\ $\lambda$, to vary, square wells
obviously make for a convenient test case of \HRT\ and, indeed, of
liquid state theories in general; consequently, a great many
simulational and theoretical efforts have been directed at this
system, and studies of its phase behavior abound
 \cite{sw:1,sw:4,sw:7,sw:10,sw:11,sw:12,sw:13,sw:14,sw:17,sw:16,sw:18,allg:14,allg:13}.  But square wells are also of interest in their own right,
serving as --- albeit somewhat crude --- models of a wide variety of
physical systems including, {\it e.~g.{}}, $\rm {}^3He$, $\rm Ne$, $\rm
Ar$,
$\rm H_2$, $\rm CO_2$, $\rm CH_4$, $\rm C_2H_6$, $n$-pentane and
$n$-butane \cite{allg:13,allg:16,allg:17} while current interest in
this potential derives mainly from the finding that square wells capture
the essential features of the interactions found in colloidal systems
\cite{coll::1,coll::2,coll::3,coll::4,coll::5}.  Yet another
motivation for this first application of \HRT\ to square wells comes
from a recent, very accurate simulation study \cite{sw:18} confirming
and quantifying the presence in the system with $\lambda=1.5$ of the
Yang-Yang (YY) anomaly expected and experimentally found for
asymmetric fluids \cite{rg:1,rg:2}.

Due to the extensive amount of data available in the literature the
more recent of which will shortly be presented later on, and in view
of some of the limitations of \HRT\ in its current formulation we
cannot expect to gain new insight into the system at hand with a level
of precision comparable to that of the more sophisticated simulation
schemes. Instead, in the present contribution our focus of interest
lies on some aspects of \HRT's numerical side, specifically on those
that are sensitive to the potential's range: indeed, as stated already
in ref.~\onlinecite{ar:4}, for a potential as pronouncedly short-ranged as
square wells some of the numerical problems should show up much more
prominently than in other systems like, {\it e.~g.{}}, the hard-core Yukawa
fluid previously considered \cite{ar:4} where they are, of course, in
principle still present but do not manifest themselves as clearly.

In accordance with the preceding remarks, another reason why
application of \HRT\ to square wells might be worthwhile lies in the
closure underlying seemingly all applications so far of \HRT\ to
simple one-component fluids with hard sphere reference part: As the
usual formulation of \HRT\ in these cases relies on an {\it ansatz}
for the two-particle direct correlation function $c_2(r)$ very much in
the spirit of Stell's {\it Lowest-Order $\gamma$-ordered
Approximation} \cite{loga1,loga2} (\LOGA) or the equivalent {\it
Optimized Random-Phase Approximation} \cite{orpa} (\ORPA) by Andersen
and Chandler, the direct correlation function can never extend to
larger $r$ values than the potential $v(r)$ itself. In particular, for
the square well potential $v^{{\rm sw}[-\epsilon,\lambda,\sigma]}(r)$ we
necessarily have $c_2(r)=0$ for $r>\lambda\,\sigma$ so that all
moments of $c_2(r)$, {\it i.~e.{}}\ $\int_{{\Bbb R}^3}{{\rm
d}}^3r\,c_2(r)\,r^n$,
$n\ge0$, exist, which is obviously at variance with the correct
behavior near the critical point \cite{allg:4}; furthermore, at
intermediate $Q$ the direct correlation function can hardly be
considered satisfactory, especially \cite{ar:4,ar:th} close to
$r=\lambda\,\sigma$. While some earlier publications
\cite{hrt:10,hrt:7,allg:7} already blamed unsatisfactory aspects of \HRT\
results
on this inadequacy of the closure, square wells should bring out
related problems of \HRT\ with the usual \LOGA/\ORPA-style closure
much more clearly, and the numerical procedure's response  should provide
us with a signature to be looked
out for in other systems, too; also, even within the \LOGA/\ORPA-style
approximation the implementation of the core condition {\it via}\
approximate ordinary differential equations (\ODE s) for the relevant
expansion coefficients easily shown to be inadequate for very
short-ranged potentials \cite{ar:th} casts some doubt on the range of
$\lambda$ values amenable to an \HRT\ treatment in the current
formulation of the theory. Determination of the admissible
$\lambda$-range, on the other hand, is particularly interesting in the
light of
refs.~\onlinecite{sw:13} and \onlinecite{allg:7} as well as in view of the
global
renormalization scheme \cite{allg:14,allg:13,allg:16,allg:17}
originally developed by White and co-workers as an extension of
Wilson's phase-space cell method \cite{wilson:phase-space-cell} to the
liquid state; it is only by combining tests internal to the theory and
comparison with data available by other means that we are able to
answer this question.

In this contribution, after a sketchy presentation of the underlying
theory itself (section~\ref{sec:theory}) and the implementation used
(section~\ref{sec:program}), in section~\ref{sec:application} we turn to
the
results of applying \HRT\ to square well systems of variable
range. After a short summary of the critical point's location as
obtained from simulation-based and other purely theoretical studies of
square wells for various values of $\lambda$ (sub-section
\ref{sec:non-hrt}) we first look into the core condition's implementation,
which provides us with a first hint regarding the range of $\lambda$
values accessible to \HRT\ in its current formulation and once more
highlights the decoupling assumption's {\it r\^ole} (sub-section
\ref{sec:cc}). The latter is also implicated in the correct choice of the
boundary condition imposed at high density as discussed, alongside the
boundary condition's location's effect, in sub-section~\ref{sec:boundary}.
--- A particularly grave aspect of \HRT's numerical side is
the stiffness of the partial differential equation (\PDE) for
close-to-critical and sub-critical temperatures
(sub-section~\ref{sec:stiff}), the vestiges  of which are evident in the
results obtained for quasi-continually varying $\lambda$ as presented
in sub-section~\ref{sec:interference}. A short summary of our findings and
conclusions ends our contribution (section~\ref{sec:bye}).

\section{The Theory}

\label{sec:theory}

The definite resource on \HRT\ is the review \cite{b:hrt:1} by the
theory's original authors, summarizing its formalism as developed in a
series of earlier publications \cite{hrt:1,hrt:2,hrt:11,hrt:10,hrt:3,hrt:4}
so that we here present only an overview of the theory used
and of its formulation, recapitulating some of our earlier findings
\cite{ar:4}; the notation employed here of course co-incides with that
of our preceding contribution \cite{ar:4}.

As mentioned already in section~\ref{sec:intro}, \HRT's mainstay is the
implementation of the suppression of long-wavelength fluctuations
characteristic of \RG\ methods by means of a renormalized potential
$v\Q$.  Thus, rather than directly going from a reference fluid the
properties of which are assumed known to the fully interacting system,
{\it i.~e.{}}\ from pair potential $v\Ref(r)$ to $v(r)=v\Ref(r)+w(r)$,
\HRT\
proceeds {\it via}\ a succession of rather artificial \cite{ar:4}
intermediate
potentials $v\Q(r)$: For every value of the cut-off wave number $Q$,
$v\Q$ is given by
\begin{displaymath}
\begin{array}{c}
v\Q(r)=v\Ref(r) + w\Q(r)
\\
\tilde w\Q(k) = \left\{\begin{array}{ccc}
\tilde w(k) &:& k>Q \\
0        &:& k<Q
\end{array}\right.
\\
\tilde w^{{\rm sw}[-\epsilon,\lambda,\sigma]}(k) =
-4\pi\,\epsilon\,
{\sin\lambda\,\sigma\,k-\lambda\,\sigma\,k\,\cos\lambda\,\sigma\,k
\over k^3}\,,
\end{array}
\end{displaymath}
where the last line specializes to the square well potential of
eq.~(\ref{eq:def:potSW}); obviously, $v\Ref$ and $v$ are recovered in the
limits $Q\to\infty$ and $Q\to0$,
\begin{displaymath}
\begin{array}{c}
v\withQ{\infty}(r) = v\Ref(r) = v^{{\rm hs}}(r)
\\
v\withQ0(r) = v(r) = v^{{\rm sw}}(r)\,,
\end{array}
\end{displaymath}
allowing \HRT\ to gradually turn on fluctuations of ever increasing
wavelength by lowering $Q$ from $\infty$ to zero (numerically
\cite{ar:4}, from $Q_\infty$ to $Q_0$); as mentioned before, criticality
(together with non-classical critical exponents) and phase separation
(with isotherms rigorously flat in the two phase region) are obtained
in the limit $Q\to0$. In this procedure it is essential to maintain
the differential picture implied by \RG\ theory and to make sure that
the transition from $Q$ to infinitesimally smaller cut-off $Q-{{\rm d}} Q$
is
continuous even in the limit $Q\to0$. The latter requirement
necessitates replacing the usual free energy $A\Q$ and two-particle
direct correlation function $c_2\Q(r)$ of the hypothetical system with
cut-off $Q$ and potential $v\Q(r)$, the ``$Q$-system'', by suitably
modified quantities, {\it viz.{}}
\begin{displaymath}
\begin{array}{c}
{\beta\A\Q\over V} = {\beta A\Q\over V}
- {\rho^2\over2}\left(\tilde\phi(0)-\tilde\phi\Q(0)\right)
+ {\rho\over2}\left(\phi(0)-\phi\Q(0)\right)
\\
\C\Q(r) = c_2\Q(r) + \phi(r) - \phi\Q(r)
\\
\phi=-\beta\,w\qquad\beta=1/\kB\,T\,,
\end{array}
\end{displaymath}
where $\rho$ is the number density of the system at hand; the higher
order correlation functions $c\Q_n(r_1,\ldots,r_n)$, $n\ge3$, are free
from such problems. (Note that all the direct correlation functions
including $\C\Q(r)$ are taken to include the ideal gas terms
\cite{b:hrt:1}.)

With this set of quantities continuous even in the limit $Q\to0$,
{\it viz.{}}\ $\A\Q$, $\C\Q$, and the $c_n\Q$, $n\ge3$, \HRT\ is derived as
a
non-terminating hierarchy of coupled \ODE s at fixed density $\rho$,
calculating the properties of the $Q$-system by treating the system at
infinitesimally higher cut-off $Q+{{\rm d}} Q$ as a reference system; of
these equations, usually only the evolution equation for $\A\Q$,
{\it viz.{}}
\begin{equation}\label{eq:dQ:A}
{{{\rm d}}\over{{\rm d}} Q}\left({\beta\A\Q\over V}\right)
=  {Q^2\over4\pi^2}\ln\left(1-{\tilde\phi(Q)\over\tilde\C\Q(Q)}\right)\,,
\end{equation}
as well as the important compressibility sum-rule
\begin{equation}\label{eq:consistency}
\tilde\C\Q(0) = - {\partial^2\over\partial\rho^2}\left({\beta\A\Q\over
V}\right)
\end{equation}
valid for any cut-off $Q$ directly enter practical calculations.

When combined with a closure on the two-particle level,
eqs.~(\ref{eq:dQ:A})
and (\ref{eq:consistency}) define a \PDE\ in the $(Q,\rho)$-plane; it is
this \PDE\ that we will concern ourselves with in the remainder of
this text. Said closure, reminiscent of \LOGA/\ORPA\ but adding one
free parameter to allow imposing thermodynamic consistency as embodied
in eq.~(\ref{eq:consistency}), is given, just as in our earlier
contribution
\cite{ar:4}, by
\begin{equation}\label{eq:closure}
\begin{array}{rl}
\C\Q(r,\rho)&{} = \phi(r,\rho)
+ \g\Q_0(\rho)\,u_0(r,\rho)
+ \K\Q(r,\rho)\\
\K\Q(r,\rho)&{} = \G\Q(r,\rho) + \c\Ref(r,\rho)\\
\G\Q(r,\rho)&{} = \sum_{n=1}^\infty\g\Q_n(\rho)\,u_n(r,\rho)\,,
\end{array}
\end{equation}
where we have generalized to density-dependent potentials. Basis
function $u_0(r,\rho)$ is chosen to coincide with \cite{ar:th}
$w(r,\rho)/\tilde
w(0,\rho)$, and the higher basis functions $u_n(r,\rho)$,
$n\ge1$, vanish outside the core; for our specific choice of basis
functions see appendix~B\ of ref.~\onlinecite{ar:4}. In order to ensure
that
both the core condition, {\it i.~e.{}}\ $g(r,\rho)=0$ for $r<\sigma(\rho)$
where
$g(r,\rho)$ is the pair distribution function, and
sum-rule~(\ref{eq:consistency}) are met it is necessary to choose the
correct set of expansion coefficients $\g_n\Q(\rho)$, $n\ge0$, at
every cut-off $Q$ and for every density $\rho$; assuming their
validity for $Q=\infty$ and adopting the short hand notations
\begin{displaymath}
\begin{array}{c}
\alpha\Q(\rho) = {\partial^3\over\partial Q\partial\rho^2}
\left({\beta\A\Q\over V}\right)\\
\hat{\cal I}\Q[\psi(k,\rho),\rho] =
\int_{{\Bbb R}^3} {{{\rm d}}^3k\over(2\pi)^3}\,{\psi(k,\rho)\over
\left(\tilde c\Q_2(k,\rho)\right)^2}
\end{array}
\end{displaymath}
(here, $\psi$ is an arbitrary function of $k$ and $\rho$), both
relations can be combined to
\begin{equation}\label{eq:cc:cInt}
\begin{array}{l}
\sum\limits_{n=1}^\infty
\hat{\cal I}\Q\left[\tilde u_j(k,\rho)\,\left(\tilde u_n(k,\rho)
-\tilde u_0(k,\rho)\,\tilde u_n(0,\rho)\right),\rho\right]
\,{\partial\g\Q_n(\rho)\over\partial Q}
\\\qquad
=
\alpha\Q(\rho)\,
\hat{\cal I}\Q\left[\tilde u_j(k,\rho)\,\tilde u_0(k,\rho),
\rho\right]
\\\qquad\quad
+ {Q^2\over2\pi^2}
\,{\tilde\phi(Q,\rho)\,\tilde u_j(Q,\rho)
\over\tilde\C\Q(Q,\rho)
\,\left(\tilde\C\Q(Q,\rho)-\tilde\phi(Q,\rho)\right)}\,,
\quad j\ge1\,;
\end{array}\\
\end{equation}
this set of equations must, of course, be truncated to a finite number
$1+{N_{\rm cc}}$ of basis functions, and it is also necessary to neglect
non-local contributions to $\partial \hat{\cal I}\Q[\psi(k,\rho), \rho] /
\partial Q$ to allow convenient evaluation at arbitrary $Q$. Both of
these approximations have been discussed at length in our previous
contribution \cite{ar:4}, and while the value of ${N_{\rm cc}}$ was found
to
strongly influence the quality of the results obtained, determination
of the $\g\Q_n(\rho)$ from eq.~(\ref{eq:cc:cInt}) and said approximation
for
the slowly converging $\hat{\cal I}$-integrals' $Q$-dependence leads to
systematic deficiencies at small $r$ in $g(r)$ as determined from the
Ornstein-Zernike relation. --- Unfortunately, for numerical reasons
\cite{ar:4} it is necessary to also adopt the so-called decoupling
assumption \cite{hrt:4}, {\it viz.{}}\ $\alpha\Q(\rho)=0$; as can easily be
seen, this is not only mathematically incompatible with thermodynamic
consistency but even suffices to decouple the \PDE\ implied by
eqs.~(\ref{eq:dQ:A}) and (\ref{eq:consistency}) to a set of unrelated \ODE
s at
fixed density only lacking thermodynamic consistency and thus unable
to predict clear phase boundaries \cite{ar:4}; furthermore, we cannot
rule out that decoupling may have a significant influence on the
solution generated \cite{ar:4}, which is particularly troublesome as the
much longer range of $u_0(r)\propto \phi(r)$ when compared with the
other basis functions was originally invoked as justification for
setting $\alpha\Q(\rho)=0$: for square wells, this assumption is
certainly even less justified than for the rather long-ranged
hard-core Yukawa system ($z=1.8/\sigma$) considered in
ref.~\onlinecite{ar:4}.

Returning to the \PDE, for the numerical implementation's benefit we, too,
adopted a re-formulation in terms of an auxiliary function
$f(Q,\rho)$ simply related to the modified free energy's derivative
with respect to $Q$. The details of the procedure leading to a \PDE\
of the form
\begin{equation}\label{eq:pde:f}
\begin{array}{c}
{\partial\over\partial Q}f(Q,\rho)
= d_{00}[ f,Q,\rho]
+ d_{01}[ f,Q,\rho]\,{\partial\over\partial\rho}f(Q,\rho)
+ d_{02}[ f,Q,\rho]\,{\partial^2\over\partial\rho^2}f(Q,\rho)\,,
\\
f(Q,\rho)\,\tilde u_0^2(Q,\rho)
= \ln\left(1-{\tilde\phi(Q,\rho)\over\tilde\C\Q(Q,\rho)}\right)
+ {\tilde\phi(Q,\rho)\over\tilde\K\Q(Q,\rho)}\,,
\end{array}
\end{equation}
and the coefficient functions $d_{0i}$ themselves can be found in
appendix~A of ref.~\onlinecite{ar:4}, {\it q.~v.{}}\
ref.~\onlinecite{ar:th}.

The above formulation~(\ref{eq:pde:f}) of the problem, of course still
coupled to the \ODE s implementing the core condition, obviously has
to be amended by initial and boundary conditions; while the former
easily follow from $\g\withQ{{Q_\infty}}_n=0$, $n\ge0$, (which is
sufficient to also determine $f({Q_\infty},\rho)$), choice of appropriate
boundary conditions is slightly more complicated: if, as is the case
in most of the calculations reported here (exceptions see below), the low
density boundary is located at
${\rho_{\rm min}}=0$, we can make use of the divergence of the ideal gas
term
$-1/\rho$ in $\tilde c\Ref_2$ to derive not only $f(Q,0)=0$ but also
$\partial f(Q,0)/\partial\rho=0$ which alone is, in principle,
sufficient to uniquely determine the solution up to arbitrarily high
density; for computational reasons, however, it is preferable to
instead only impose vanishing $f$ at ${\rho_{\rm min}}$ and to supply an
approximate condition for calculating $f$ at ${\rho_{\rm max}}$. Among the
candidates for the constraint to be imposed upon the solution at
${\rho_{\rm max}}$ in addition to the core condition there are two we
should
mention here: Starting with ref.~\onlinecite{hrt:3}, the so-called
\ORPA-condition, {\it viz.{}}\ $\g\Q_0({\rho_{\rm max}})=0$, has been used
extensively;
it should, however, be noted that this condition is incompatible with
both thermodynamic consistency and the decoupling assumption
\cite{ar:4}.  An alternative first considered in our previous report
\cite{ar:4} is the decoupling assumption $\alpha\Q({\rho_{\rm max}})=0$
itself;
of course, this condition is still incompatible with the
compressibility sum-rule (\ref{eq:consistency}) but this is less of a
problem at a boundary where partial derivatives with respect to $\rho$
cannot be evaluated anyway. Another option (not pursued in this
contribution) is to give up the core condition altogether, retaining
only the lowest basis function $u_0$ in the closure~(\ref{eq:closure}) and
thus effectively replacing eq.~(\ref{eq:cc:cInt}) by
eq.~(\ref{eq:consistency});
this has the added advantage of mathematical consistency while still
retaining the structure of a \PDE\ so important for obtaining clear
phase boundaries \cite{ar:4}, {\it v.~s.{}}

It is one of \HRT's main achievements to allow calculating a fluid's
binodal (coinciding with the spinodal in three dimensions \cite{hrt:8})
without resorting to Maxwell constructions \cite{hrt:8}, for subcritical
temperatures yielding isotherms rigorously flat in density intervals
the boundaries of which are readily identified with the coexisting
densities $\rho_v$ and $\rho_l$. Thus, as the isothermal
compressibility $\kappa_T$ of the fully interacting system, readily
found to be proportional to $\exp(f-(\tilde\phi(0)/\tilde\K\withQ0))-1$
(cf.~appendix~A of ref.~\onlinecite{ar:4}), diverges in the two-phase
region, so must the auxiliary function $f(Q,\rho)$ in the limit
$Q\to0$. As a direct consequence of this, the transition from the
modified free energy $\A\Q(\rho)$ to $f(Q,\rho)$ is not only
computationally convenient but also allows us to follow the isothermal
compressibility's build-up much more easily; even more importantly, a
simple analysis \cite{ar:4,ar:th} of the behavior of the \PDE's
coefficients for large $f(Q,\rho)$ readily characterizes the \PDE\ as
stiff: for any density $\rho\in[\rho_v,\rho_l]$ and close to $Q=0$,
the true solution $f(Q,\rho)$ oscillates rapidly on a $Q$-scale of the
order of $\exp(-f)$, with both an upper bound on the oscillations'
amplitudes and $f$'s average slope growing like $1/Q$ --- needless to
say that this behavior cannot be reproduced numerically ({\it v.~i.{}}\
sub-section~\ref{sec:stiff}; {\it q.~v.{}}\ ref.~\onlinecite{ar:4}). Note,
however, that it is
not an artifact of the re-writing of the \PDE\ in the form (\ref{eq:pde:f})
but rather a problem inherent to \HRT\ itself in a formulation based
upon~eq.~(\ref{eq:closure}) \cite{ar:th}.

\section{Numerical procedure}

\label{sec:program}

The numerical study of \HRT\ for square well systems of varying range
parameter $\lambda$ in section~\ref{sec:application} has only become
feasible due to our recent re-implementation\cite{ar5:fn:1} 
of this theory, discussed at length in refs.~\onlinecite{ar:4} and
\onlinecite{ar:th}; we will make use of results obtained with this program
exclusively. From a practical point of view, our software provides a
means of solving a finite-difference approximation to the
\PDE~(\ref{eq:pde:f}) in an iterated full-approximation scheme, imposing
boundary
conditions at densities ${\rho_{\rm min}}$ and ${\rho_{\rm max}}$ as well
as initial
conditions at $Q=Q_\infty$, generating a solution for $Q$ as low as
$Q_0$ while ensuring numerical soundness of every step by employing a
number of criteria. The pivotal parameter governing all of the
numerics  is a small quantity denoted
$\epsNum$ characteristic of the maximum admissible relative error
introduced in a single step in the $-Q$ direction; due to the
paramount importance of derivatives with respect to $\rho$, $\epsNum$
is strictly related to the coarseness of the density grid.

The only exception to the general strategy of ensuring a numerical
quality of $\epsNum$ at every step in the calculation is the choice of
step sizes $\Delta Q$ in the $-Q$ direction, at least for sub-critical
and close-to-critical temperatures: indeed, in that part of the
$(Q,\rho)$-plane where the divergence of the isothermal
compressibility builds up, the \PDE's stiffness ({\it v.~s.{}}) renders
fixed-precision arithmetic and relative errors bounded by $\epsNum$
incompatible. Consequently, for the calculations reported below we
resort to step sizes $\Delta Q$ pre-determined in a way analogous to
that employed in earlier applications \cite{hrt:10,ar:4}; still,
monitoring and assessing suitable components of the solution vector in
terms of $\epsNum$ as described in sub-section~III~E\ of
ref.~\onlinecite{ar:4} may yield a wealth of information on the numerical
process and its evolution.

Most of the calculations reported here have been performed on an
equispaced density grid of ${N_{\rho}}=100$ density intervals spanning the
range from ${\rho_{\rm min}}=0$ to ${\rho_{\rm max}}=1/\sigma^3$,
corresponding to a
value of $\epsNum=10^{-2}$; ${N_{\rm cc}}$ was usually set to~7; and the
pre-determined step sizes started from $\Delta Q=-10^{-2}/\sigma$ at
${Q_\infty}=80/\sigma$, plunging to a mere $-5\cdot10^{-6}/\sigma$ when
approaching ${Q_0}=10^{-4}/\sigma$. --- When locating the binodal {\it
via}\
the divergence of the isothermal compressibility $\kappa_T$ we did not
require an actual overflow to occur but instead looked for a
$\kappa_T$-ratio at neighbouring densities exceeding $10^4$, which is
a rather reliable indicator for the binodal's location as $\kappa_T$
typically jumps by far less than two or by at least some twenty orders
of magnitude within one $\Delta\rho$; the reported values for $\rho_v$
and $\rho_l$ are the mid-points of the density intervals so found.  In
principle this allows us to locate the coexisting densities and the
critical temperature and density to arbitrary precision, even though
the computational cost rises sharply with falling $\epsNum$.

\section{Application to square wells}

\label{sec:application}

As mentioned before, much of the motivation for applying \HRT\ in the
formulation outlined in section~\ref{sec:theory} to the simple square well
model potential is based upon various observations indicating possible
limitations of this approximate formulation of \HRT\ for short-ranged
potentials. A case in point is the recent work of Caccamo {\it et al.{}}\
\cite{allg:7} entirely devoted to several thermodynamically consistent
theories' ability to deal with narrow hard-core Yukawa systems; sure
enough, in the case of \HRT\ the shortcomings of the \LOGA/\ORPA-style
closure (\ref{eq:closure}) and, presumably, of the accompanying decoupling
assumption underlying the core condition's implementation {\it via}\
eq.~(\ref{eq:cc:cInt}) were manifest already in refs.~\onlinecite{hrt:10}
and
\onlinecite{hrt:7} and recently confirmed by us \cite{ar:4}, {\it q.~v.{}}\
ref.~\onlinecite{ar:th}.

Of course, any of the problems discussed below only relate to \HRT\
when implemented along the lines of sections~\ref{sec:theory} and
\ref{sec:program} and not to \HRT\ proper; however, for reasons discussed
in
ref.~\onlinecite{ar:4}, alternative formulations almost certainly render
the
numerics far more demanding and open up a whole new suite of problems
regarding the numerical implementation's soundness, especially when
performing Fourier transformations of cut-off affected functions
\cite{ar:th}.

In the following sub-sections we will complement the discussion of
ref.~\onlinecite{ar:4} by further investigation into the numerical nature
of
\HRT; before that, however, it seems pertinent to re-iterate some of
the points raised in that publication as far as they concern the
reasoning to be put forward in the following. In particular, according
to section~IV of ref.~\onlinecite{ar:4}, for the numerical results to be
meaningful the coexisting densities $\rho_v$ and $\rho_l$ must
maintain a separation of at least several density grid spacings
$\Delta\rho$ from the boundaries at ${\rho_{\rm min}}$ and ${\rho_{\rm
max}}$;
consequently, $\beta$ should never exceed some maximum value,
$\beta<{\beta_{\rm max}}$, and for the systems considered here and in
ref.~\onlinecite{ar:4} and for the typical choices for ${\rho_{\rm min}}$
and
${\rho_{\rm max}}$ the binodal's proximity to the low density boundary
renders
${\beta_{\rm max}}$ largely density grid- and $\epsNum$-independent. ---
Not to
be confused with ${\beta_{\rm max}}$ is the lowest temperature
$\kB/{\beta_{{\rm max},\tt\#}}$ numerically accessible to the program with
pre-determined step sizes: this is the temperature below which the
program of section~\ref{sec:program} never reaches $Q\approx Q_0$ or
produces abnormal results; note that ${\beta_{{\rm max},\tt\#}}$ may be
larger or
smaller than ${\beta_{\rm max}}$, depending on the chosen combination of
physical potential, approximations in the formulation used (the
boundary conditions in particular), and the choice of parameters
affecting the numerical work.

Regarding the implementation of the core condition as sketched in
section~\ref{sec:theory}, the main conclusion of ref.~\onlinecite{ar:4} was
that a
minimum of ${N_{\rm cc}}=7$ basis functions in addition to $u_0$ were
necessary for acceptable results despite residual defects of $g(r)$
close to the origin; a short discussion of the core condition's
slightly different {\it r\^ole} for square wells will be presented
below (sub-section~\ref{sec:cc}).

\begin{figure}\epsfbox{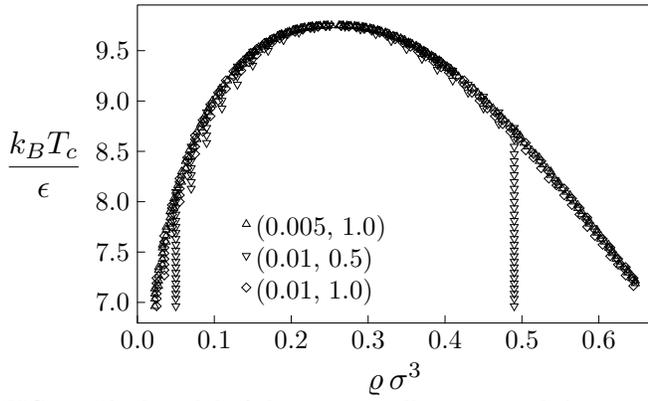}
\caption{The binodal of the square well system with $\lambda=3$ as
obtained for different values of $(\epsNum,{\rho_{\rm max}})$, cf.~the
discussion in sub-section~\ref{sec:boundary}. Note that for this rather
long-ranged system the critical point's location is virtually
unaffected by variation of these parameters. Also, imposing the
boundary condition at ${\rho_{\rm max}}=0.5/\sigma^3$ clearly induces a
shift
in $\rho_v$ to higher and, to a lesser degree, in $\rho_l$ to lower
values even well above the temperature where $\rho_l$ gets close to
${\rho_{\rm max}}$, which is readily interpreted as an effect brought about
by
stiffness; the results for ${\rho_{\rm max}}=0.5/\sigma^3$ and
$\epsNum=0.005$
(not shown in the plot) do not differ much from those with the same
${\rho_{\rm max}}$ and $\epsNum=0.01$ except in the binodal's vapor
branch's
shift being somewhat smaller.}
\label{fig:rhomax}
\end{figure}

The critical density $\rho_c$ predicted by \HRT, it should be noted,
is virtually always in reasonable agreement with literature data as
shortly presented in sub-section~\ref{sec:non-hrt}; indeed, \HRT\ is even
able to reproduce the marked rise in $\rho_c$ predicted by
refs.~\onlinecite{sw:1}, \onlinecite{sw:13}, and \onlinecite{sw:17} for
$\lambda\to1+$ as
opposed to the rigorously constant value in ref.~\onlinecite{sw:14}. Due to
the satisfactory $\rho_c$-values obtained numerically we will
henceforth exclude $\rho_c$ from the discussion; for a demonstration
of both $\rho_c$'s insensitivity to variation of parameters of the
numerical procedure and the quantitative agreement with the data of
sub-section~\ref{sec:non-hrt} see figures~\ref{fig:rhomax} and
\ref{fig:crit:lambda}.

In this context it may be of interest that the \HRT\ estimate for the
critical density presents no difficulties for the hard-core Yukawa
fluid considered in ref.~\onlinecite{ar:4}, either, nor is there any
mention of
such difficulties in any of the other publications on this topic that
we are aware of; indeed, the theory's numerical problems primarily lie
in the solution's small-$Q$ behavior for close-to-critical and
sub-critical temperatures on the one hand and the use of mutually
incompatible assumptions prompted by the need to employ decoupling
without giving up thermodynamic consistency on the other hand.  Both
of these aspects of \HRT\ pertain to different parts of the
$(Q,\rho)$-plane, located close to the high density boundary for the
{\it r\^ole} mathematical inconsistencies play and at not too large
$Q$ and $\rho\sim\rho_c$ for the pathological behavior related to
coexistence; they will be discussed in sub-sections~\ref{sec:boundary}
and~\ref{sec:stiff}, respectively, and their vestiges will also be seen in
the results of applying \HRT\ in the formulation of
section~\ref{sec:theory} to square wells of quasi-continually varying range
$\lambda\in\,] 1,3.6]$ in sub-section~\ref{sec:interference}.

\subsection{Non-\HRT\ results for square wells}

\label{sec:non-hrt}

For comparison purposes we have collected in tables \ref{tb:ref.sim} and
\ref{tb:ref.theo} the critical temperatures of various square well systems
as obtained from simulations (table~\ref{tb:ref.sim}) or by purely
theoretical means (table~\ref{tb:ref.theo}); the data included have been
published within the last decade.

Of the simulation based results included in table~\ref{tb:ref.sim}, only
those of ref.~\onlinecite{sw:12} for $\lambda\in\{1.25, 1.375, 1.5, 1.75,
2\}$ have been obtained by molecular dynamics (MD); most of the other
simulation studies rely on one or the other variant of the Monte Carlo
(MC) method: Among these, the Gibbs ensemble MC (GEMC) calculations of
ref.~\onlinecite{sw:1} set out to determine critical exponents, $\beta$ in
particular; that work's finding of $\beta \sim 1/2$ for $\lambda=2$ as
opposed to the expected $\beta\sim1/3$ found for $\lambda$ up to 1.75
prompted re-examination of the square well fluid with $\lambda=2$ by
GEMC augmented by finite-size scaling (FSS) techniques \cite{sw:4},
refuting the mean field value for the effective exponent. ---
Especially in the critical regime, grand canonical MC (GCMC)
simulations incorporating histogram reweighting and FSS offer some
advantage over GEMC due to the latter's restriction to fixed
temperature; such an approach has been applied to square wells with
$\lambda=1.5$ and~3 in ref.~\onlinecite{sw:7}; a more elaborate GCMC scheme
not biased towards the Ising universality class and taking into
account the YY anomaly has recently been applied to \cite{sw:18}
$\lambda=1.5$, {\it v.~s.{}}{} Yet another method goes under the name of
thermodynamic- or temperature-and-density-scaling MC (TDSMC); it was
applied to the case of $\lambda=1.5$ and analyzed in terms of an
effective Hamiltonian in refs.~\onlinecite{sw:10} and \onlinecite{sw:11}.
--- Also
included in table~\ref{tb:ref.sim} are the results of
ref.~\onlinecite{sw:13}, employing
an MC scheme modified to take advantage of a speed-up possible by
combining simulation data with an analytical {\it ansatz} for the
chemical potential; the efficiency of this approach originally devised
to study phase separation allows a large number of systems to be
considered. (The error bounds given for these ``modified MC'' results
in table~\ref{tb:ref.sim} have been obtained from the different results
displayed in ref.~\onlinecite{sw:13} for different parameter settings.)

The theoretical predictions for the critical temperature listed in
table~\ref{tb:ref.theo} comprise a second-order analytic perturbation
theory
\cite{sw:17} (APT2) applicable to $1<\lambda\le2$ and claimed accurate
for $\lambda\ge1.4$ as well as the hard-sphere van der Waals (HSvdW)
equation of state \cite{sw:14}. In addition, though not listed in
table~\ref{tb:ref.theo}, we have utilized the non-square-well-specific
Okumura-Yonezawa (OY) estimate \cite{allg:15} for $\beta_c$,
primarily as a starting value when looking for the critical
temperature in our \HRT\ calculations; for
$v^{{\rm sw}[-\epsilon,\lambda,\sigma]}$, the OY prediction is
$\kB\,T_c/\epsilon=0.203\,(2\pi/3)\,\lambda^3-0.273$.

\subsection{The core condition}

\label{sec:cc}

Ever since application of \HRT\ to continuous fluids started, the
implementation of the core condition has been a major issue, probably
motivating adoption of the closure~(\ref{eq:closure}) and variants thereof
for non-hard-sphere reference systems \cite{hrt:17} despite its known
deficiencies in the first place; indeed, it is no coincidence that
several studies \cite{hrt:3,hrt:6,hrt:19,hrt:13} primarily concerned
with the \RG\ aspect of the theory chose to completely eliminate the
core condition. When applying \HRT\ as a regular liquid state theory,
on the other hand, this is not an option: too great is the effect this may
have on
both correlation functions and phase behavior \cite{ar:4}. From
table~\ref{tb:Ncc} where we compile the critical temperature
$T_c=1/\kB\,\beta_c$
for various square well potentials as functions of the number ${N_{\rm
cc}}+1$
of basis functions in the closure~(\ref{eq:closure}), just as in
ref.~\onlinecite{ar:4} we find virtually constant critical temperatures for
$1\le{N_{\rm cc}}\le4$; on the other hand, the amount of variation seen
upon further
increasing ${N_{\rm cc}}$ strongly depends on $\lambda$, which immediately
carries over to the pair distribution function $g\withQo(r,\rho)$ and
its compatibility with the core condition: For $\lambda=3$, the
longest ranged potential considered in table~\ref{tb:Ncc},
$g\withQo(r,\rho)=0$, $r<\sigma$, holds reasonably well except very
close to $r=0$ even for ${N_{\rm cc}}=1$; when increasing the number of
basis
functions all the way to ${N_{\rm cc}}=10$, the pair distribution function
has
to be corrected for very small $r$ only, yielding a $\vert
g\withQo(r,\rho)\vert$ that remains bounded by some $10 ^{-2}$ of the
contact value $g\withQo(\sigma+,\rho)$ for all $r<\sigma$; the
corresponding small change in $g\withQo(r,\rho)$ and
$\C\withQo(r,\rho)$ is reflected in the near-constant predictions for
$\beta_c$ evident from table~\ref{tb:Ncc}. Similarly, for $\lambda=1.5$ and
$\lambda=2$ and within the ${N_{\rm cc}}$-range considered, the
implementation
of the core condition does not convincingly improve except for
supercritical temperatures and intermediate densities; this time,
however, the pair distribution functions remain far from compatible
with the core condition even for ${N_{\rm cc}}=10$, and neither $\beta_c$
nor
$g\withQo(r,\rho)$ itself nor, for that matter, the final values of
the \LOGA/\ORPA\ expansion coefficients $\g\withQo_n(\rho)$ indicate
that the expansion (\ref{eq:closure}) for $\tilde\C\Q(k,\rho)$ might be
close
to convergence. But if the quality of $g\withQo(r,\rho)$ improves only
little if at all, the remaining deficiencies are probably to be blamed
on the approximation for the poorly convergent integrals' derivative
with respect to $Q$ mentioned earlier (cf.\ eq.~(12)\ of
ref.~\onlinecite{ar:4}) rather than on an insufficient number of basis
functions; on the other hand, even though the decoupling assumption cannot
directly affect the pair distribution function's compliance with the
core condition, the approximation of neglecting the non-local term in
$\partial \hat{\cal I}\Q \left[\psi(k,\rho), \rho\right] / \partial Q$ is
on the same level as that of setting $\alpha\Q(\rho)=0$, as was
stressed by the authors of ref.~\onlinecite{hrt:4} upon jointly introducing
these two assumptions. Thus, combining the above findings regarding
the core condition with the analogous analysis of sub-section~IV of
ref.~\onlinecite{ar:4} and with that contribution's investigation into the
decoupling assumption's possible effect (cf.\ fig.~2\ of
ref.~\onlinecite{ar:4}) we are led to the conclusion that decoupling poses
certainly no less a problem here than for the hard-core Yukawa
potential studied there.

\subsection{High-density boundary condition}

\label{sec:boundary}

Numerically, there are two ways for the implementation of
section~\ref{sec:program} to fail to reach $Q=Q_0$, both, of course, easily
detected by
the ``monitoring'' variant of our code (cf.\ sub-section~III~E\
of ref.~\onlinecite{ar:4}): due to the solution's pathological behavior
wherever
$f(Q,\rho)$ is large (cf.\ sub-section~\ref{sec:stiff}), or because of
inappropriate boundary conditions at high density. As for the latter
--- an issue intimately linked to the decoupling assumption ---, the
immediate reason for the program's failure is a near-discontinuity in
the numerical solution close to the boundary: For the moment setting
aside the decoupling assumption and other approximations, in the
application of \HRT\ with the closure (\ref{eq:closure}) at any point
$(Q,\rho)$ in the interior of the \PDE's domain the core condition
uniquely determines the $\g_n\Q(\rho)$, $n\ge1$, for given
$\g_0\Q(\rho)$; this expansion coefficient is then determined by
imposing thermodynamic consistency as embodied in the compressibility
sum-rule (\ref{eq:consistency}). At a boundary, {\it i.~e.{}}\ for
$\rho\in\{{\rho_{\rm min}},{\rho_{\rm max}}\}$, however, the second density
derivative
cannot be evaluated reliably so that some other condition must be
imposed; in the calculations reported here (with the obvious exception
of those for fig.~\ref{fig:C:interference}) we always choose ${\rho_{\rm
min}}=0$
so that the divergence of the ideal gas term in $\tilde c_2\Q$ provides
$f(Q,\rho)=0$ as a convenient and unproblematic boundary
condition. For $\rho={\rho_{\rm max}}$, on the other hand, we are in
principle free to use
any suitable approximation for the structural and thermodynamic
properties of the $Q$-system and to calculate $f(Q,{\rho_{\rm max}})$ from
said
approximation, thereby providing the necessary boundary condition for
the \PDE~(\ref{eq:pde:f}); but for practical reasons it is desirable to use
the same \LOGA/\ORPA-form for the $Q$-system's direct correlation
function at ${\rho_{\rm max}}$ as in the rest of the problem's domain so
that,
in particular, the \LOGA/\ORPA\ prescription $\g\Q_0({\rho_{\rm max}})=0$
is a
natural choice of boundary condition.  In general, however, due to the
\PDE's diffusion-like character any condition imposed at ${\rho_{\rm max}}$
that is incompatible with the solution for $\rho<{\rho_{\rm max}}$ by
necessity
induces a corresponding near-discontinuity in $f(Q,\rho)$ close to the
boundary; within the framework of a finite difference scheme this is
reflected in a mismatch of $f(Q,{\rho_{\rm max}})$ and the solution at
densities close by, {\it i.~e.{}}\ $f(Q,{\rho_{\rm max}}-i\,\Delta\rho)$
for small
$i\ge1$, and the mismatch's severity may serve as a direct measure for
the inappropriateness of the boundary condition at ${\rho_{\rm max}}$ in
relation to the approximations applied at densities in
$\left]{\rho_{\rm min}}, {\rho_{\rm max}}\right[$.

On the other hand, the numerics become intractable unless we adopt the
decoupling
assumption, and the only way to consistently use $\alpha\Q(\rho)=0$
without abandoning the core condition is to decouple the \HRT-\PDE\ to
a set of \ODE s at fixed density only \cite{ar:4}; this, unfortunately,
removes all traces of thermodynamic consistency from the equations and
thereby precludes obtaining clear phase boundaries \cite{ar:4}. It is
therefore necessary to restrict decoupling to the implementation of
the core condition only while retaining the structure of a \PDE\
together with the compressibility sum rule (\ref{eq:consistency}) despite
the latter's incompatibility with decoupling. Thus, for
${\rho_{\rm min}}<\rho<{\rho_{\rm max}}$, both $\tilde\C\Q(0,\rho) = -
\partial^2
\left({\beta \A\Q / V}\right) / \partial\rho^2$ and $\alpha\Q(\rho)=0$
are used for different parts of the problem; at ${\rho_{\rm max}}$,
however,
again any approximation allowing calculation of $f(Q,{\rho_{\rm max}})$ may
be
used, so that it is tempting to once again resort to the
\LOGA/\ORPA-condition of vanishing $\g\Q_0({\rho_{\rm max}})$ or variants
thereof.

But due to the decoupling assumption's possibly large effect, any
boundary condition that does not incorporate $\alpha\Q({\rho_{\rm max}})=0$
---
and bear in mind that $\g\Q_0({\rho_{\rm max}})$ and $\alpha\Q({\rho_{\rm
max}})$ cannot
both vanish at the same time for generic cut-off $Q$ --- will once
again incur a fatally large mismatch; if, however, we must resort to
decoupling anyway, it seems preferable to consistently apply it for
the boundary condition rather than to inconsistently combine it with a
condition alien to the theory; also, though the mismatches' magnitudes
from imposing $\alpha\Q({\rho_{\rm max}})=0$ alone or from mixing it with
the
\LOGA/\ORPA\ condition $\g\Q_0({\rho_{\rm max}})=0$ generally do not differ
much as long as the \PDE's stiffness does not play a {\it r\^ole}
({\it e.~g.{}}, for $\lambda=1.5$, as long as we restrict ourselves to
$Q\sim8/\sigma$ or higher, or to $\beta\ll\beta_c$), the former
approach fares better than the other one more often than not. It is
only in this sense, {\it i.~e.{}}\ presupposing a \LOGA/\ORPA-like {\it
ansatz}
even at ${\rho_{\rm max}}$ and application of decoupling in the
implementation
of the core condition according to eq.~(\ref{eq:cc:cInt}) at all $\rho$,
that the results are largely independent of the choice of boundary
condition as claimed, for $\beta<\beta_c$, in ref.~\onlinecite{hrt:4}.

In the numerical work we find that such a mismatch is present whenever
the calculation proceeds {\it via}\ mathematically inconsistent or
conflicting approximations; in the case of square wells with their
comparatively short potential range, however, the problems are much
more severe than in other systems so that ${\beta_{{\rm max},\tt\#}}$ is
rather
small and even drops below $\beta_c$ for most of the $\lambda$
interval from 1 to 2 (cf.~sub-section~\ref{sec:interference}).  Restricting
ourselves to $\beta<{\beta_{{\rm max},\tt\#}}$ and $Q=Q_0$, the
mismatch is typically reflected in an increase by one order of
magnitude in the three-point finite-difference estimate of, {\it e.~g.{}},
$\vert\partial^2 f(Q_0,\rho) / \partial\rho^2\vert$ right at the
boundary over the near-constant values at slightly lower densities;
apart from a positive correlation with $\epsNum$, the mismatch's
severity is qualitatively unaffected by a change in parameters of the
numerical procedure or the choice and location of the boundary
condition (with the above provisions).

Another effect worth mentioning in connection with the boundaries is
the influence their locations, {\it viz.{}}\ ${\rho_{\rm min}}$ and
${\rho_{\rm max}}$, may
have. The basic mechanism and its implications for the coexisting
densities were already mentioned in the opening remarks of this
section; here we only want to point out that the non-criticality
enforced by the boundary conditions not only may unduely distort the
binodal predicted by \HRT\ as demonstrated in fig.~\ref{fig:rhomax}, very
small ${\rho_{\rm max}}$ may also allow one to reach $Q=Q_0$ at higher
$\beta$,
thus effectively raising ${\beta_{{\rm max},\tt\#}}$ while lowering
${\beta_{\rm max}}$. ---
Sometimes, however, the expectation of the binodal keeping a
separation from the boundary of several $\Delta\rho$ at least does not
hold, and a preposterous two-phase region appears very close to
${\rho_{\rm max}}$ or, very rarely, close to ${\rho_{\rm min}}$; {\it
e.~g.{}}, for $\lambda=1.88$ and
$\beta=0.392/\epsilon$ the equations can be solved all the way down to
$Q=Q_0=10^{-4}/\sigma$, predicting an unrealistic two-phase region
extending from $0.845(5)/\sigma^3$ to $0.995(5)/\sigma^3$. This behavior
turns out to come from the interplay of the mismatch at ${\rho_{\rm max}}$
and
the numerical treatment of the stiffness of the \PDE.

\subsection{The region of large $f(Q,\rho)$}

\label{sec:stiff}

For subcritical temperatures the \HRT-\PDE's true solution's erratic
behavior in that part of the $(Q,\rho)$-plane where $f(Q,\rho)$ is
large and the isothermal compressibility's divergence is built up
(cf.~section~\ref{sec:theory}, {\it q.~v.{}}~ref.~\onlinecite{ar:th})
obviously eludes reliable
numerical realization; in particular, while $\epsNum$ still
characterizes the level of accuracy in auxiliary calculations, the
same can no longer be true for the accuracy of the \PDE's
discretization as this would require step sizes $\Delta Q$ so small as
to cause floating point underflow upon evaluating, {\it e.~g.{}},
$Q-(Q-\Delta
Q)$, thus rendering finite differences numerically insignificant.

Consequently, in this respect we have to give up our strategy of
controlling the numerical procedure so as to locally ensure a quality
of $\epsNum$ at least, turning to pre-determined step sizes \cite{ar:4}
$\Delta Q$ in addition to fixed $\Delta\rho$, to which similar
concerns apply \cite{ar:th}; on such a coarse mesh of $(Q,\rho)$-points
underlying the finite difference scheme, however, the true solution
cannot even be represented adequately, and the numerical approximation
for $f(Q,\rho)$ obtained from the \PDE's discretization with these far
too large step sizes cannot be trusted to faithfully represent even
the average behavior of $f(Q,\rho)$.

This inadequacy of the step sizes is reflected in various
peculiarities of the solution vector obtained in the numerical
procedure; indeed, when monitoring the evolution of $f(Q,\rho)$ and
the core condition coefficients $\g\Q_n(\rho)$, our code readily
detects the plummeting step sizes necessary and signals the
incompatibility of the behavior seen with the assumption of smoothness
underlying finite difference schemes. Another telltale sign is
iterated corrector steps' failure to converge when $f(Q,\rho)$ is
large: even though implicit schemes like the one we employ \cite{ar:th}
are the standard treatment for stiff systems, the rapid growth of the
oscillations' amplitudes renders the finite difference equations
themselves unstable under iteration; only when resigning on any
control of the numerical error and refraining from iterations of the
corrector step do the step sizes $\Delta Q$ chosen allow one to force
advancing $Q$ all the way to $Q_0$ in remarkably many cases. Also,
comparison of $f(Q_0,\rho)$ as obtained with different sets of step
sizes $\Delta Q$ reveals that, for $\rho_v<\rho<\rho_l$, the evolution
of $f(Q,\rho)$ seen numerically is driven by the number and size of
$Q$-steps only and certainly does not correspond to an average over
oscillations \cite{ar:th}; the same mechanism is also responsible for a
small $\Delta Q$-dependence of the critical temperature $\beta_c$. ---
By the same token, due to the $d_{01}$- and $d_{02}$-terms in
eq.~(\ref{eq:pde:f}), the \PDE's stiffness and the related problems have a
direct
bearing on the solution outside the coexistence region even if the
numerical predictions there turn out rather insensitive to variation
of parameters of the numerical procedure; in particular, we expect a
gradual but non-negligible distortion (in addition to the effects of
numerical differentiation close to the near discontinuity) of the
binodal, increasing with falling temperature.

\subsection{\HRT\ results for square wells}

\label{sec:interference}

In the light of the preceding exposition as well as of the discussion
in ref.~\onlinecite{ar:4} it may at first seem surprising that \HRT\ in the
formulation of section~\ref{sec:theory} has a record of being well
applicable to a variety of systems (cf.~section\ \ref{sec:intro}); also,
as we shall see in a moment, even for square wells, a system expected
to be particularly vulnerable to the problems just outlined, we find
reasonable estimates of the critical points' locations for a wide
range of $\lambda$-values. Still, the mechanisms sketched in
subsections~\ref{sec:boundary} and~\ref{sec:stiff} as well as the
difficulties
presented in ref.~\onlinecite{ar:4} remain and manifest themselves
numerically in a
number ways.

For a first orientation, let us look at the results summarized in
figs.~\ref{fig:crit:lambda} and~\ref{fig:crit:lambda:details} where the
critical temperature $T_c$ and density $\rho_c$ are shown as functions
of $\lambda$; the underlying calculations have been obtained with
$\epsNum=10^{-2}$, imposing decoupling in a consistent way at
${\rho_{\rm max}}=1/\sigma^3$ and with ${N_{\rm cc}}+1=7+1$ basis functions
in the
expansion~(\ref{eq:closure}) of the \LOGA/\ORPA-function $\G\Q$. With the
exception of some spurious results at $\lambda\sim1.1$, whereever
$\beta_c<{\beta_{{\rm max},\tt\#}}$ the critical temperature in general
compares
quite favorably with the data of tables~\ref{tb:ref.sim}
and~\ref{tb:ref.theo}; from the calculations we have performed for a large
number
of systems in the range $1<\lambda\le3.6$ and ignoring some isolated
results, a critical point is found for $1.06\le\lambda\le1.24$, for
$1.45\le\lambda\le1.53$, and for $\lambda\ge1.939$; calculations with
${N_{\rm cc}}=5$ yield analogous results \cite{ar:th}, with
$\beta_c<{\beta_{{\rm max},\tt\#}}$ in a somewhat larger part of the
parameter
range, {\it viz.{}}\ for $1.09\le\lambda\le1.58$ and for $\lambda\ge1.896$,
but will not be considered in the following in view of the
considerations of sub-section~\ref{sec:cc} and of other defects that turn
out to be larger than for ${N_{\rm cc}}=7$.

For the moment setting aside the data for $\lambda<1.939$, \HRT's
predictions for the critical temperature are generally found to be in
satisfactory agreement with the $\beta_c(\lambda)$-curve expected from
the simulation-based and theoretical results presented in sub-section
\ref{sec:non-hrt}. Embedded into this regular overall behavior of
$\beta_c$ as a function of $\lambda$, however, we find a number  of
depressions and elevations
of $\beta_c$, some of which cannot be seen on the scale of the
plot~\ref{fig:crit:lambda} but from the numeric results only \cite{ar:th};
others,
however, are so strong as to render the critical temperature a
non-monotonic function of $\lambda$, which is certainly not expected
on the grounds of the literature presented in
sub-section~\ref{sec:non-hrt}, the data of
refs.~\onlinecite{sw:13,sw:14,sw:17} in particular.

In the light of sub-section~\ref{sec:stiff} it is of course tempting to
simply attribute this behavior to the difficulties previously
discussed, especially since the critical point is located in the
region of large $f(Q_0,\rho)$ by definition; the peculiar distribution
of $\lambda$-values affected, however, suggests that these problems of
the numerical procedure are triggered by a special mechanism. Indeed,
a closer look at the core condition function $\tilde\C\Q(Q,\rho)$ for
fixed density $\rho$ reveals, for every single one of the $\lambda$
values implicated that we checked, that the combination of terms
pertaining to $w(r)$ or $v^{{\rm hs}}(r)$ alone (of ranges
$\lambda\,\sigma$
and $\sigma$, respectively) regularly and quite frequently reduces the
amplitude of this function's swings about the ideal gas value of
$-1/\rho$; the same happens only occasionally for $\lambda$-values
removed from these irregularities so that it is, in fact, possible to
quite reliably determine whether or not a given $\lambda$ is affected
from a plot of $\tilde\C\Q(Q,\rho)$ for $\rho\sim\rho_c$ alone as
illustrated in fig.~\ref{fig:C:interference}. It will come as no surprise
that most of the irregularities occur when $\lambda$, the ratio of the
two characteristic lengths present in the model, is close to a simple
fraction: among the shifts in $T_c$ most obvious are those where
$\lambda$ is close to $2$ (cf.\ fig.~\ref{fig:crit:lambda:details}),
$2{1\over4}$, $2{1\over7}$, $2{1\over9}$ and $2{1\over12}$, and in
retrospect it seems justified to also include the small parameter
range around $\lambda=1{1\over2}$ in this list, {\it v.~i.{}}; the effect
is
less obvious from fig.~\ref{fig:crit:lambda} but still discernible at
$2{1\over2}$, $2{1\over3}$ and $2{2\over3}$, whereas for $2{1\over4}$
and $2{3\over4}$ it is so small as to make the plot of
$\beta_c(\lambda)$ appear smooth while the irregularities are still
evident from the numerical values; also note that, once again,
$\rho_c$ is hardly affected.

All these observations seem to indicate that indeed it is the
interplay of the two different length scales and the resulting partial
oppression of a significant portion of the oscillations of
$\tilde\C\Q(Q,\rho)$ that cause the discrepancy of \HRT\ and literature
results for the critical temperature around certain $\lambda$ values
even though a smooth interpolation of \HRT's predictions from
$\lambda$ values nearby is well compatible with the data presented in
sub-section~\ref{sec:non-hrt}. Even though we currently cannot pinpoint
the precise mechanism by which this unphysical behavior of \HRT\
arises and, in particular, cannot distinguish between the closure's
inadequacy and the \PDE's stiffness as the main culprit --- though the
latter is certainly implicated to some degree 
---, two conclusions may be drawn quite safely: for one, as long as we
stay clear of values of $\lambda\simgt2$ that are close to simple
fractions, or restrict ourselves to $\lambda\simgt2.7$ where the
effects are rather small, we can probably trust the numerical results
--- with the {\it caveats} of ref.~\onlinecite{ar:4} and
sub-sections~\ref{sec:boundary} and~\ref{sec:stiff} --- to the same degree
of confidence as
those obtained for the hard-core Yukawa system in ref.~\onlinecite{ar:4}.
And
secondly, it is only in the presence of discontinuities in the
potential that certain lengths feature prominently in the relevant
functions' Fourier transforms and can so give rise to problems of the
kind outlined above; consequently, as long as we confine ourselves to
continuous $w(r,\rho)$, which still includes most of the potentials
popular in liquid state physics, the unphysical shifts in $\beta_c$
seen for certain parameter combinations are likely not an issue,
whereas the same problems are expected to resurface, {\it e.~g.{}}, for the
multi-step potential also defined in sub-section\ III~B\ of
ref.~\onlinecite{ar:4}.

Another lesson to be drawn from the findings presented here as well as
in ref.~\onlinecite{ar:4} is that, as a general rule, conclusions should
never be
drawn from isolated results alone; it is only through the combination
and meticulous scrutiny of a set of related calculations that
meaningful information can be extracted from \HRT\ calculations: due
to the problems related to the implementation of the core condition,
to the nature and location of the boundary conditions, and to the
\PDE's stiffness, any single calculation must be considered as of
uncertain standing. As an example \cite{ar:th}, the analogue  of
fig.~\ref{fig:rhomax} for $\lambda=1.5$ shows  a considerably larger
variation in the binodal and the
critical point's location, which is consistent with the above
conclusions regarding the reason for ${\beta_{{\rm max},\tt\#}}$ rising
above
$\beta_c$ in a narrow region around this $\lambda$ value, whereas any
one of the phase boundaries found in itself is a perfectly plausible
candidate for the ``true'' \HRT\ binodal.

\section{Conclusion}

\label{sec:bye}

In conjunction with the findings of ref.~\onlinecite{ar:4}, the discussion
of
section~\ref{sec:application} provides quite coherent a picture of \HRT's
numerical side as well as of some peculiarities encountered for square
wells. Most prominently, we see a marked dependence of the quality of
the results on the potential's range, confirming the trend of
decreasing accuracy for narrower potentials reported \cite{allg:7}
for the hard-core Yukawa fluid; it has long been accepted
\cite{hrt:10,hrt:7,allg:7} that the simplistic but computationally
convenient
\cite{ar:4} closure (\ref{eq:closure}) has a part in this, and an improved
closure has recently been proposed \cite{cond-mat:0109311}. Still, as
far as numerical application of \HRT\ is concerned the closure cannot
be discussed without reference to the decoupling assumption and to the
approximate implementation of the core condition {\it via}\ \ODE s coupled
to the \HRT-\PDE; and while the former has been found problematic both
for square wells (present contribution) and for the hard-core Yukawa
fluid considered in ref.~\onlinecite{ar:4} and should probably not be
trusted
easily for any system, the severity of the difficulties brought about
by the simplified treatment of the core condition sensitively depends
on the potential type and parameters chosen: for the continuous and
rather long ranged Yukawa potential with $z=1.8/\sigma$,
$g\withQo(r,\rho)$ can be made sufficiently small within the core, and
the square well fluid with $\lambda=3$ fares equally well at least;
from the discussion of sub-section~\ref{sec:cc} and the data of
table~\ref{tb:Ncc}, however, it becomes apparent that smaller $\lambda$ ---
we have
looked at $\lambda=1.5$ and $\lambda=2$ in particular --- incurs
substantial problems, with residual defects in the pair distribution
functions attributed to the ill-justified approximation of neglecting
some slowly converging integrals \cite{ar:4,ar:th} in the core
condition.

But table~\ref{tb:Ncc} demonstrates not only the $\lambda$-dependence of
the
results' sensitivity to the number ${N_{\rm cc}} + 1$ of basis functions
retained in the truncated eq.~(\ref{eq:cc:cInt}) when varying ${N_{\rm
cc}}$ in the
range $0\le{N_{\rm cc}}\le10$: while the virtually constant critical
temperature predicted for $\lambda=3$ seems trustworthy and is,
indeed, well compatible with simulation results (cf.\
table~\ref{tb:ref.sim}), the amount of variation in $\beta_c$ for
$\lambda=1.5$ and,
to a much lesser degree, for $\lambda=2$ precludes accurate
determination of the critical temperature; this is a first indication
that the theory might be able to handle square wells with $\lambda=3$
quite reliably whereas problems cannot be denied for $\lambda=2$, and
$\lambda=1.5$ seems largely out of reach for \HRT\ in the present
formulation.  This is confirmed by the results obtained by
quasi-continuous variation of $\lambda$ in the range $1<\lambda\le3.6$
as shown, for ${N_{\rm cc}}=7$, in fig.~\ref{fig:crit:lambda}: the critical
point
is accessible only in part of this parameter range, and not only the
critical temperature at fixed $\lambda$ but also the boundaries of the
$\lambda$-intervals where \HRT\ is able to reach temperatures as low
as $T_c$ strongly depend on ${N_{\rm cc}}$ (cf.\
sub-section~\ref{sec:interference}).

Applicability of \HRT\ to only a restricted $\lambda$-range is, of
course, again related to the pronounced short-rangedness of the square
well potential; to explain it, however, we have to invoke not only the
\LOGA/\ORPA-style closure and the approximate implementation of the
core condition but also the other difficulties encountered in the
numerical procedure as highlighted in this and our preceding
contribution, {\it viz.{}}\ the decoupling assumption
(ref.~\onlinecite{ar:4}), inappropriate
boundary conditions (sub-section~\ref{sec:boundary}) and, most
importantly, the \PDE's stiffness for thermodynamic states of high
compressibility (sub-section~\ref{sec:stiff}). All of these are, in
principle, always present to some degree in numerical applications of
\HRT; it may prove valuable that sub-sections~\ref{sec:boundary}
and~\ref{sec:stiff} provide distinct signatures readily detected by the
implementation of section~\ref{sec:program} that might be looked out for
in more advanced applications of the theory, too. --- Related to these
difficulties is a peculiar effect specific to square wells
(sub-section~\ref{sec:interference}): close to certain $\lambda$ values,
simple fractions in particular, we see shifts in the critical
temperature that render \HRT's predictions much less compatible with
simulations and other theoretical descriptions of the square well
fluid than would be expected from the results obtained for $\lambda$
values close by; the mechanism for triggering these local distortions
of the $\beta_c(\lambda)$-curve is illuminated at least to the point
of linking it to the presence of a discontinuity in the potential's
perturbational part. All in all, the numerical evidence as well as
comparison with literature data suggest that the formulation of \HRT\
sketched in section~\ref{sec:theory} is well able to deal with square
wells and to locate their critical points to reasonable accuracy for
$\lambda\simgt2$ as long as certain values are avoided, or else for
$\lambda\simgt2.7$.

\section{Acknowledgments}

The authors gratefully acknowledge support
from {\it\"Osterreichischer Forschungsfonds} under project number
P13062-TPH.

\begin{table}
\begin{tabular}{|c|clc|}
$\lambda$&$k_B\,T_c(\lambda)/\epsilon$&method&\\
\hline
\hline
1.05&0.3751(1)&mod.~MC \cite{sw:13}&\\
\hline
1.1&0.4912(4)&mod.~MC \cite{sw:13}&\\
\hline
1.15&0.5942(35)&mod.~MC \cite{sw:13}&\\
\hline
1.2&0.692(1)&mod.~MC \cite{sw:13}&\\
\hline
1.250&0.764(4)&GEMC \cite{sw:1}&\\
\hline
1.25&0.78&MD \cite{sw:12}&\\
&0.7880(6)&mod.~MC \cite{sw:13}&\\
\hline
1.3&0.8857(7)&mod.~MC \cite{sw:13}&\\
\hline
1.375&0.974(10)&GEMC \cite{sw:1}&\\
&1.01&MD \cite{sw:12}&\\
\hline
1.4&1.076(8)&mod.~MC \cite{sw:13}&\\
\hline
1.5&1.2179(3)&GCMC(YY) \cite{sw:18}&\\
&1.2180(2)&GCMC \cite{sw:7}&\\
&1.219(8)&GEMC \cite{sw:1}&\\
&1.222&TDSMC \cite{sw:10,sw:11}&\\
&1.226&TDSMC \cite{sw:10,sw:11}&\\
&1.246(5)&TDSMC \cite{sw:10,sw:11}&\\
&1.27&MD \cite{sw:12}&\\
&1.302(8)&mod.~MC \cite{sw:13}&\\
\hline
1.65&1.645(5)&mod.~MC \cite{sw:13}&\\
\hline
1.75&1.79&MD \cite{sw:12}&\\
&1.811(13)&GEMC \cite{sw:1}&\\
\hline
1.8&2.062(8)&mod.~MC \cite{sw:13}&\\
\hline
2&2.61&MD \cite{sw:12}&\\
&2.648(14)&GEMC+FSS \cite{sw:4}&\\
&2.666(85)&GEMC+FSS \cite{sw:4}&\\
&2.678(27)&GEMC+FSS \cite{sw:4}&\\
&2.6821(8)&GEMC+FSS \cite{sw:4}&\\
&2.684(51)&GEMC+FSS \cite{sw:4}&\\
&2.721(89)&GEMC+FSS \cite{sw:4}&\\
&2.730(14)&GEMC+FSS \cite{sw:4}&\\
&2.764(23)&GEMC \cite{sw:1}&\\
&2.778(7)&mod.~MC \cite{sw:13}&\\
\hline
2.2&3.80(1)&mod.~MC \cite{sw:13}&\\
\hline
2.4&5.08(2)&mod.~MC \cite{sw:13}&\\
\hline
3&9.87(1)&GCMC \cite{sw:7}&\\
\end{tabular}
\caption{The critical temperature $T_c$ of square well systems for
various values of $\lambda$ as predicted by simulations and
simulation-based theoretical analyses, and the corresponding
references. The acronyms used for labeling the method employed in
obtaining these results are given in sub-section~\ref{sec:non-hrt} of the
text.}
\label{tb:ref.sim}
\end{table}

\begin{table}
\begin{tabular}{|c|clc|}
$\lambda$&$k_B\,T_c(\lambda)/\epsilon$&method&\\
\hline
\hline
1.125&0.587&APT2 \cite{sw:17}&\\
\hline
1.25&0.751&HSvdW \cite{sw:14}&\\
&0.850&APT2 \cite{sw:17}&\\
\hline
1.375&0.978&HSvdW \cite{sw:14}&\\
&1.08&APT2 \cite{sw:17}&\\
\hline
1.5&1.249&HSvdW \cite{sw:14}&\\
&1.33&APT2 \cite{sw:17}&\\
\hline
1.625&1.61&APT2 \cite{sw:17}&\\
\hline
1.75&1.859&HSvdW \cite{sw:14}&\\
&1.93&APT2 \cite{sw:17}&\\
\hline
1.85&2.23&APT2 \cite{sw:17}&\\
\hline
2&2.506&HSvdW \cite{sw:14}&\\
&2.79&APT2 \cite{sw:17}&\\
\end{tabular}
\caption{The critical temperature $T_c$ of square well systems for
various values of $\lambda$ as predicted by purely theoretical means,
and the corresponding references. The acronyms used for labeling the
method employed in obtaining these results are given in
sub-section~\ref{sec:non-hrt} of the text.}
\label{tb:ref.theo}
\end{table}

\begin{table}
\begin{tabular}{|c|cccc|}
${N_{\rm cc}}$&
$k_B\,T_c(\lambda=1.5)/\epsilon$&
$k_B\,T_c(\lambda=2)/\epsilon$&
$k_B\,T_c(\lambda=3)/\epsilon$&\\
\hline
--&
1.209437(035)&
2.660946(132)&
9.891032(298)&\\
\hline
1&
1.190663(034)&
2.682489(105)&
9.899937(478)&\\
2&
1.203326(035)&
2.686289(105)&
9.900894(478)&\\
3&
1.200152(035)&
2.686078(105)&
9.900894(478)&\\
4&
1.197136(034)&
2.685655(105)&
9.900894(478)&\\
5&
1.287443(040)&
2.527365(093)&
9.737080(462)&\\
6&
1.098329(029)&
2.742404(110)&
9.822071(471)&\\
7&
0.984757(047)&
2.914763(124)&
9.867502(475)&\\
8&
1.070878(027)&
2.744830(110)&
9.773324(466)&\\
9&
1.216333(036)&
2.749695(110)&
9.887510(477)&\\
10&
1.207583(035)&
2.937591(126)&
9.748203(464)&\\
\end{tabular}
\caption{The dependence of the critical temperature of square well
systems on the number ${N_{\rm cc}}+1$ of basis functions retained in
eqs.~(\ref{eq:closure}) and (\ref{eq:cc:cInt}) for various values of
$\lambda$. For
${N_{\rm cc}}>0$, the decoupling assumption was imposed as high density
boundary condition, whereas the \LOGA/\ORPA-condition
$\g\Q_0({\rho_{\rm max}})=0$ served the same purpose for ${N_{\rm cc}}=0$;
other
parameters were chosen as indicated in section~\ref{sec:program}.}
\label{tb:Ncc}
\end{table}

\begin{figure}\epsfbox{figure2.ps}
\caption{The critical temperature $T_c$ (dots in upper panel) and critical
density $\rho_c$ (bars in lower panel) of square well systems for
$\lambda$ ranging from close to unity up to 3.6 as obtained from
calculations with parameters chosen as indicated in
section~\ref{sec:program}; also included are the non-\HRT\ predictions
listed in
tables~\ref{tb:ref.sim} and~\ref{tb:ref.theo}, labeled by the acronyms
introduced in sub-section~\ref{sec:non-hrt} and already used in those
tables. The ticks on the top border of the figure's frame indicate the
$\lambda$ values considered; of the 200-odd systems we looked at,
${\beta_{{\rm max},\tt\#}}$ exceeded $\beta_c$ only in the $\lambda$ ranges
indicated in sub-section~\ref{sec:interference} or for some isolated
$\lambda$ values outside those ranges. The three boxes in the upper
panel indicate the parameter ranges displayed at larger scale in
fig.~\ref{fig:crit:lambda:details}.  In the lower panel, the bars show the
coexisting densities found according to the prescriptions of
section~\ref{sec:program} for the highest-temperature sub-critical
isotherm calculated in locating the critical temperature, which
explains the apparent differences in $\rho_c$'s accuracy; the smallest
$\rho_c$ intervals shown coincide with the spacing
$\Delta\rho=10^{-2}/\sigma^3$ of the density grid.}
\label{fig:crit:lambda}
\end{figure}

\begin{figure}\epsfbox{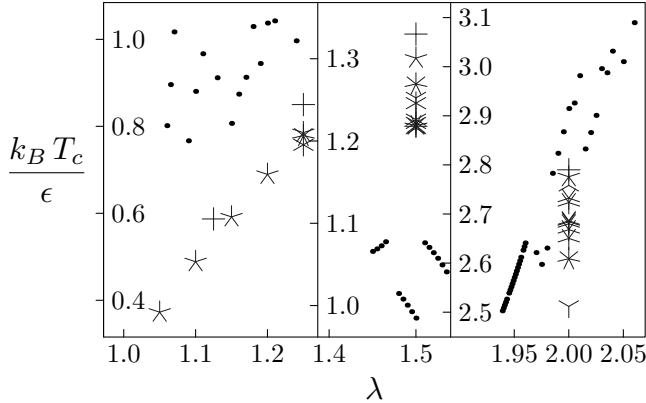}
\caption{The critical temperature data of fig.~\ref{fig:crit:lambda} for
values of the square well range parameter $\lambda$ close to 1.1, 1.5,
and~2 at larger scale; the symbols coincide with those used in
fig.~\ref{fig:crit:lambda}.}
\label{fig:crit:lambda:details}
\end{figure}

\begin{figure}\epsfbox{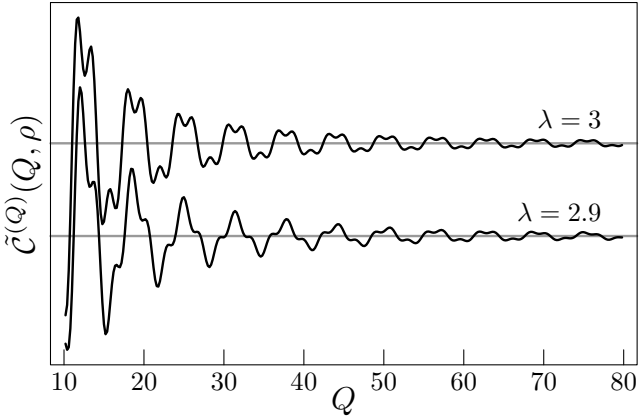}
\caption{The core-condition function $\tilde\C\Q(Q,\rho)$ for
$\rho=0.3/\sigma^3$, $\beta=0.2/\epsilon$ and for two different ranges
$\lambda$ of the square well potential, on arbitrary scales; the
horizontal lines correspond to the ideal gas value $-1/\rho$. Note
that, for $\lambda=3$ (upper curve), the peak of every single one of
the function's swings is partially reduced, which is the case less
than half the time --- and at rather high $Q$ only --- for
$\lambda=2.9$ (lower curve). We have excluded the data for
$Q<10/\sigma$ so that the effects of the \PDE's stiffness are still
negligible; the underlying calculations have been performed by solving
the \ODE s corresponding to consistent application of the decoupling
assumption at the density indicated.}
\label{fig:C:interference}
\end{figure}

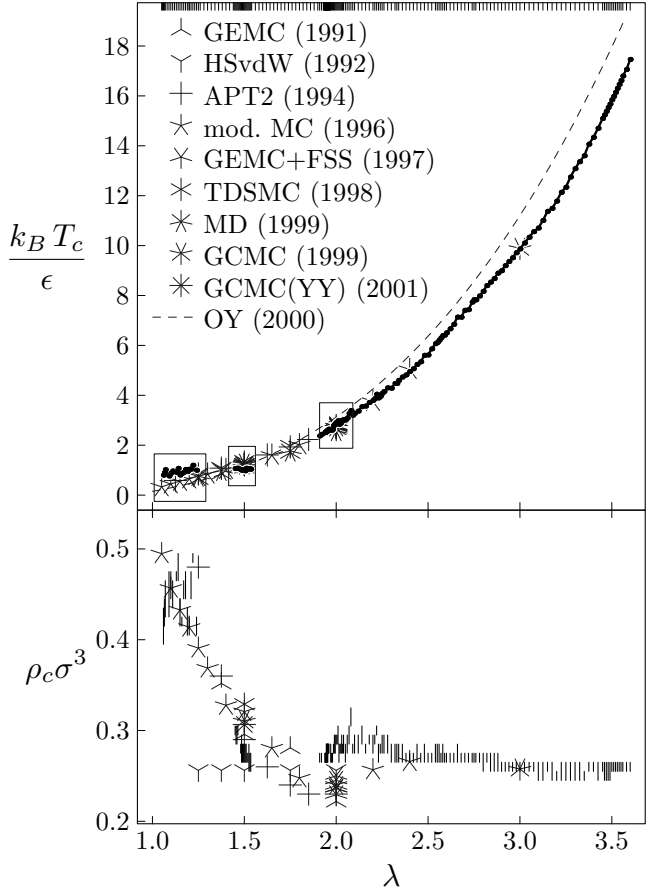
\end{document}